# Socratic Mind: Impact of a Novel GenAI-Powered Assessment Tool on Student Learning and Higher-Order Thinking

Jeonghyun Lee[a], Jui-Tse Hung[a], Meryem Yilmaz Soylu[a], Diana Popescu[a], Christopher Zhang Cui[b], Gayane Grigoryan[a], David A Joyner[a], Stephen W Harmon[a]

[a] Georgia Institute of Technology, USA
[b] University of California, San Diego, USA

**Abstract**

This study examines the impact of Socratic Mind, a Generative Artificial Intelligence (GenAI)-powered formative assessment tool that employs Socratic questioning to support student learning in a large, fully online undergraduate-level computing course. Employing a quasi-experimental, mixed-methods design, we investigated participants' engagement patterns, the influence of user experience on engagement, and impacts on both perceived and actual learning outcomes. Data were collected from the system logs, surveys on user experience and perceived engagement and learning gains, student reflections, and course performance data. Results indicated that participants consistently reported high levels of affective, behavioral, and cognitive engagement and these were strongly linked to positive user experiences and perceived learning outcomes. Quantitative analysis further revealed that students who engaged with the GenAI tool experienced significant gains in their quiz scores compared to those who did not, particularly benefiting students with lower baseline achievement. Additionally, thematic analysis of qualitative feedback revealed substantial perceived improvements in higher-order thinking skills, including problem-solving, critical thinking, and self-reflection. Our findings highlight the promise of AI-mediated dialogue in fostering deeper engagement and higher-order cognitive skills. As higher education institutions expand GenAI integration in curriculum, this dialogic, GenAI-powered assessment tool can offer a scalable strategy to promote students' meaningful learning outcomes.

*Keywords*: generative AI, formative assessment, higher-order thinking, student engagement, learning outcome

## 1. Introduction

Artificial Intelligence (AI) is profoundly reshaping educational assessment practices by automating routine evaluation tasks and delivering personalized learning experiences. With the rise of generative AI (GenAI) tools like ChatGPT, assessments can now offer real-time feedback and adapt to individual learner profiles (Ansari et al., 2024; Heil et al., 2025). However, current AI-based systems often emphasize lower-level cognitive skills, such as factual recall and procedural tasks, rather than fostering higher-order thinking abilities such as analysis, synthesis, and critical reasoning (Foung et al., 2024; Roll & Wylie, 2016). Addressing these limitations is crucial for developing AI tools that support deeper learning.

Socratic questioning, a pedagogical method rooted in structured inquiry and dialogue, offers a framework for cultivating higher-order thinking. Research has demonstrated that engaging with Socratic dialogue enhances students' ability to question assumptions, articulate reasoning, and reflect on their understanding (Chin & Osborne, 2008). Yet, traditional methods of implementing Socratic dialogue require extensive instructor involvement, limiting scalability. GenAI platforms integrating Socratic questioning hold promise for addressing this challenge by delivering scalable and adaptive assessments (Authors, 2024).

Learner engagement—cognitive, emotional, and behavioral—plays a critical role in the successful use of AI tools (Heung & Chiu, 2025). Design features and User Experiences (UX) significantly influence the depth and quality of learner interaction (Chi & Wylie, 2014; D'Mello & Graesser, 2023; Sajja et al., 2025). Students' varying levels of AI literacy, prior experience, and confidence further impact how they engage with these tools (Guettala et al., 2024). So far though, existing studies report mixed findings on GenAI's impact on learning outcomes, which requires further empirical research (Bauer et al., 2025).

This study introduces Socratic Mind, a novel GenAI-powered assessment tool that employs dynamic Socratic questioning through adaptive interactions. It aims to understand students' experiences with the tool, measure its effectiveness in improving learning outcomes, and explore its impact on students' higher-order thinking skills, including problem-solving, critical thinking, and self-reflection.

## 2. Literature Review

### 2.1. Integration of GenAI-Assisted Assessment Tools in Higher Education

Rapidly advancing AI technology is transforming assessment in higher education by automating traditionally manual tasks, such as grading, personalized feedback, and evaluating learner progress (Luo, 2024; Weng et al., 2024). GenAI is capable of producing human-like feedback that closely resembles human responses by leveraging the capabilities of large language models (LLMs) (Foung et al., 2024; Weng et al., 2024). GenAI-empowered assessment systems dynamically adapt to individual learners by utilizing sophisticated algorithms, diagnosing misconceptions, and suggesting personalized learning pathways (Heil et al., 2025; Zhu & Zhang, 2025). These adaptive systems provide timely, responsive, and personalized feedback, effectively supporting formative and summative assessment processes (Yan et al., 2024; Yilmaz & Yilmaz, 2023).

Recent research has indicated the growing potential of integrating GenAI tools into educational assessment design. For instance, Foung et al. (2024) showed how GenAI can support language learning assessments by allowing students to strategically use and evaluate online language tools, such as ChatGPT and conventional AI platforms like Grammarly, across different stages of academic writing. In the context of holistic essay scoring, Tate et al. (2024) found that the current iterations of ChatGPT performed comparably to human raters, supporting its use in low-stakes or formative assessment settings. Beyond writing assessments, AI-powered platforms such as AI Hub have enabled multi-turn conversational interactions where students actively engage with domain-specific problem-solving tasks (Sajja et al., 2025).

Despite the initial promising usage of GenAI in assessment practices, researchers have raised concerns, including ethical issues and institutional policies related to academic integrity, such as student cheating, originality of work, as well as privacy and potential biases in the GenAI models (Heil et al., 2025; Luo, 2024). Addressing these challenges requires proactive efforts from educators, who need to redesign traditional assessment approaches to adopt more innovative GenAI-integrated methods that not only leverage AI's capabilities but also facilitate students' long-term growth and self-directed learning (Weng et al., 2024). Integrating GenAI into curricula necessitates rethinking how to define and measure learning outcomes that reflect complex thinking skills such as critical thinking, creativity, synthesis of new ideas, and deep reasoning (Ansari et al., 2024). Our study addresses this gap by examining the application and impact of an AI-based formative self-assessment tool designed to foster students' deep cognitive engagement.

## 2.2. Socratic Questioning in Learning and Assessment: Potential Application of GenAI

Originating from classical philosophy, Socratic questioning has evolved significantly within modern pedagogical frameworks. Rooted in constructivist and metacognitive theories, Socratic questioning promotes active, reflective, and dialogic learning crucial for assessing and developing students' critical thinking skills (Paul & Elder, 2006). This method engages learners through structured, inquiry-based dialogue designed to challenge assumptions, encourage evidence-based reasoning, and deepen conceptual understanding (Paul & Elder, 2006; Chin & Osborne, 2008).

Educational research consistently demonstrates the effectiveness of Socratic questioning in fostering critical and reflective thinking. Regular exposure to Socratic dialogue enhances students' analytical and argumentative capacities and promotes deeper exploration of content (Paul & Elder, 2006; Yang et al., 2005). Furthermore, this approach supports metacognitive skill development, enabling students to monitor their understanding, identify misconceptions, and clearly articulate reasoning processes (Prince, 2004). Integrating Socratic dialogues within structured learning environments significantly boosts cognitive engagement and deepens conceptual understanding (Paul & Elder, 2006).

Despite these documented benefits, traditional implementations of Socratic questioning often require intensive instructor facilitation, which limits scalability, especially in large-scale or distance-learning educational contexts. In response, emerging research has investigated the potential of integrating GenAI technologies as a robust opportunity to effectively scale and automate Socratic pedagogical strategies (Authors, 2024; Fakour & Imani, 2025; Favero et al., 2024). While the research findings demonstrate the promising role of GenAI technologies in extending the benefits of Socratic inquiry to a broader range of students through scalable, data-driven approaches, further empirical studies are needed to their effectiveness across various learning settings, which became the motivation of our study.

## 2.3. Student Engagement and Learning with GenAI-Assisted Educational Tools

Student engagement is a critical factor determining the effectiveness of AI-driven educational tools, which encompasses emotional, behavioral, and cognitive dimensions, from students' affective responses like enjoyment to active participation and deep intellectual effort (Chi & Wylie, 2014; D'Mello & Graesser, 2023). Prior research suggests that UX with such tools, including ease of use, sense of usefulness, and responsiveness, substantially influence these engagement dimensions by shaping how students interact with educational content and perceive their value (Sajja et al., 2025). Personalization further enhances engagement, as AI tools tailor content and feedback to individual learner needs, thereby increasing motivation, self-efficacy, and perceived relevance (Kong & Yang, 2024; Siddiqui et al., 2025; Yilmaz & Yilmaz, 2023).

Individual differences such as prior technology experience, AI literacy, self-efficacy and confidence, and perceived relevance also affect how learners engage with AI systems. Learners with higher AI literacy often report fewer barriers to adoption and deeper interactions, while those with limited familiarity may face frustration, reducing engagement quality (Authors, 2025; Foung et al., 2024). Bauer et al. (2025) stressed that both students and teachers need foundational AI literacy skills to critically evaluate AI-generated outputs to prevent unproductive reliance on AI assistance. Similarly, Marrone et al. (2024) demonstrated that effective student-AI teamwork depends on students' ability to build trust and clarify AI's role and limitations.

Empirical evidence indicates positive impacts of student engagement with GenAI-assisted educational tools on learning outcomes (Dai et al., 2024; Gu & Yan, 2024). For example, Gu & Yan (2024) conducted a meta-analysis specifically examining GenAI interventions and reported a large positive effect size (g = 0.683) on student academic performance, supporting GenAI's potential to enhance learning outcomes across disciplines and educational levels. Previous research has also shown GenAI tools' potential to reduce cognitive load by automating structured tasks and freeing cognitive resources for more complex cognitive processes, such as creative problem-solving and critical thinking (Alavi & Westerman, 2023). Systems integrating dialogue-based interactions, reflective questioning, and adaptive scaffolding can also significantly enhance learning depth and transferability, supporting the development of durable skills essential for lifelong learning (Cole et al., 2021; Kong & Yang, 2024)

Despite these strengths, caution is warranted when interpreting claims about GenAI's learning benefits. Weidlich et al. (2025) argue that many ChatGPT-focused studies suffer from methodological limitations, including unclear treatment descriptions, poorly matched control groups, and outcome measures that fail to capture durable learning, which makes effect sizes difficult to interpret. Similar concerns arise when AI tools are used in ways that promote "metacognitive laziness" (Fan et al., 2025; Kosmyna et al., 2025; Stadler et al., 2024). For instance, Kosmyna et al.'s (2025) EEG study found that long-term reliance on LLMs was associated with reduced neural connectivity, originality, and perceived ownership of work, signaling potential erosion of deeper learning skills. To fully realize the educational potential of GenAI, future tools must balance adaptive personalization with structured, cognitively supportive designs that encourage deep, reflective engagement and mitigate overreliance (Bauer et al., 2025; Stadler et al., 2024).

## 2.4. The Current Study

To address the challenges with traditional instruction and assessment activities, researchers have developed Socratic Mind, a scalable GenAI-powered assessment tool designed to simulate Socratic dialogue through dynamic and personalized interactions. Built with LLMs and automatic speech recognition, this AI platform engages learners in real-time questioning that challenges them to verbalize their reasoning, clarify assumptions, and refine their understanding (Authors, 2024).

Grounded in principles of active learning (Freeman et al., 2014; Prince, 2004), this tool prioritizes spontaneous articulation of knowledge, encouraging students to externalize thought processes, uncover misconceptions, and practice real-time argumentation. The tool supports various types of tasks, including multi-turn questioning, short answer for recall or basic reasoning, role-play through simulated conversations, and structured debate between student and AI. Overall, activities in this AI platform aim to foster students' ability to express complex ideas, which are also considered as important durable skills for lifelong learners in the modern society (Cole et al., 2021).

This AI tool works by prompting students with instructor-defined opening questions and adapting its follow-up inquiries based on the students' spoken or written responses until the system detects satisfactory conceptual clarity. Students can move on to the next part or end the conversation by clicking a designated button. Upon completion, the platform delivers immediate, personalized feedback highlighting strengths and areas for improvement. Instructors play an important role in configuring question prompts, desired answers, and evaluation rubrics, as well as identifying common misconceptions to guide the tool's adaptive questioning logic. It also includes AI-powered question design features that help instructors craft questions based on uploaded instructional materials such as lecture notes and slides.

The current study aims to examine students' perceived learning experiences with the GenAI-powered assessment tool and evaluate the effectiveness of the tool on improving their learning outcomes by using a quasi-experimental research design. Specifically, this study is guided by three research questions, as follows:

Q1: What are the patterns of students' engagement with using the AI-powered assessment tool, and how does UX influence the levels of engagement?
Q2: Does the use of the tool make a positive effect on students' learning outcomes, both perceived and actual?
Q3: How does the use of the tool influence students' higher-order thinking skills?

## 3. Method

### 3.1. Participants and Settings

The final dataset includes 173 students who were enrolled in a 16-week long online asynchronous course, CS1301 Introduction to Computing, during Spring 2025. As an undergraduate-level course targeted mostly for freshmen, CS1301 was offered by a technology-focused research university in the U.S. and was delivered through the Canvas learning management system and edX platform. This course covers the foundations of computing, and it does not presuppose any advance knowledge of computer science. The course consists of five units with the following topics: Unit 1 (computing), Unit 2 (Procedural Programming), Unit 3 (Control Structures), Unit 4 (Data Structures), and Unit 5 (Advanced Topics: Object-Oriented Programming and Computer Algorithms).

The participants consisted of 124 (72%) female and 49 (28%) male students. Mean age was 19.4 years old. Of the total, 36% self-identified as Asian, 36% White, 9% African American or Black, 4% Hispanic or Latino origin, and 15% Other. Most of the participants reported English as a native language (85%). Regarding the general frequency of using generative AI tools such as ChatGPT, 33% of the total participants reported a few times a week or daily, 35% a few times a month, 19% less than a few times a month, and 13% never or almost never. In terms of the usage of AI in their everyday lives, 13% of the total reported a few times a week or daily, 26% a few times a month, 25% less than once a month, and 36% never. On the other hand, regarding the frequency of using AI for school, 19% of the total reported a few times a week or daily, 49% a few times a month, 17% less than a month, and 15% never. This suggests that the majority of the students (85%) report some level of AI use for school, highlighting increasing integration with AI technologies in educational settings.

### 3.2. Procedures

This study adopted a quasi-experimental research design. A total of 579 students enrolled in CS1301 were randomly assigned to complete the AI-based assignments for either Unit 3 or Unit 4. These optional assignments, accessed via a web link, offered students the opportunity to earn 3 extra credit points. Participants could complete the assignments at their own pace, with the flexibility to type or speak their responses. Each assignment was aligned with the corresponding unit quiz topics, with both assignments designed to be parallel in structure and assessment format. Each included four major topics from the unit, and students answered one to three adaptive questions per topic, resulting in 11 questions for Unit 3 and 9 for Unit 4. These questions required students to demonstrate conceptual understanding through role-playing, identifying and fixing code errors, and applying programming principles to real-world scenarios in Python (see Figure 1 for an example).

At the end of the semester, students were invited to complete an online survey regarding their experience with the GenAI-powered tool. Participation in both the assignment and survey determined inclusion in the final dataset. Of the 579 enrolled students, 147 students never attempted the extra-credit AI assignment, and an additional 89 students submitted their responses after the assignment deadline. Among the remaining 293 students, 122 either did not complete or only partially completed the survey. This yielded a final sample of 171 students: 81 students in the Unit 3 Group and 90 students in the Unit 4 Group, each having completed both the assignment and post-assignment survey on time.

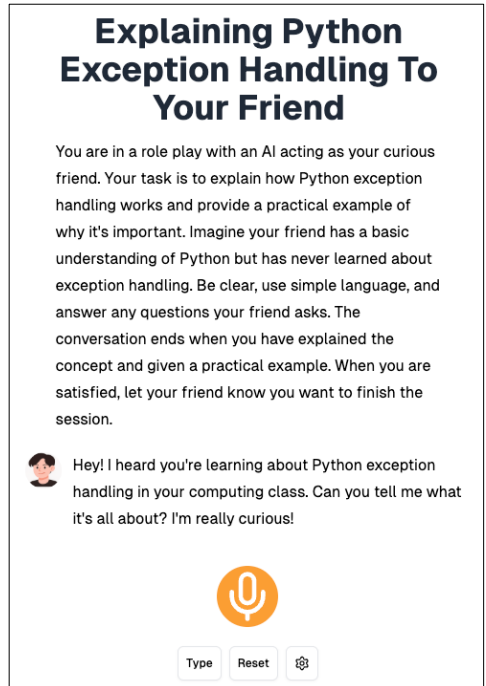

**Fig. 1** A sample role-play question from Unit 3 Assignment

### 3.3. Measures

*3.3.1. Engagement and Learning Scale*

This scale was designed to measure participants' perceived levels of engagement and learning outcomes after interacting with the GenAI tool. It consisted of three engagement sub-scales (17 closed-ended items): affective (6 items; e.g., *I enjoyed the back-and-forth dialogue with the tool*), behavioral (4 items; e.g., *I spent more time engaging with the content due to the prompts from the tool*) and cognitive engagement (7 items; e.g., *I found myself reflecting on my prior knowledge during the questioning sessions*). Also, the scale included 6 closed-ended items about perceived learning (e.g., *I believe I learned more effectively through this assignment compared to traditional methods*) and 4 short open-ended questions about their higher-order thinking skills.

For the closed-ended questions, participants were asked to read each given statement and indicate their agreement on a five-point Likert scale (1 = Strongly Disagree to 5 = Strongly Agree). The open-ended questions asked participants to describe any changes they had noticed in their problem-solving, critical thinking, self-reflection, and verbal communication skills. In addition, the engagement and learning sub-scales showed robust internal reliability scores, indicated by Cronbach's Alpha, with $\alpha = .90$ for affective engagement; $\alpha = .76$ for behavioral engagement; $\alpha = .85$ for cognitive engagement; and $\alpha = .86$ for perceived learning.

*3.3.2. UX Survey*

A 10-item UX survey was designed to capture participants' experiences with the GenAI-powered tool in three broad aspects: usability (5 items) and perceived learning effectiveness (5 items). The usability subscale focused on how much the tool supports user autonomy (e.g., *The AI tool's interface is user-friendly and easy to follow*). The perceived learning effectiveness subscale assessed participant perceived learning gains during the interaction with the tool (e.g., *The AI tool helped enhance my understanding of the topic*). Participants were asked to rate the given statements on a 5-point Likert scale (1 = Strongly Disagree to 5 = Strongly Agree). Internal consistency was strong across all three subscales ($\alpha = .95$. for usability and $\alpha = .97$. for perceived learning effectiveness). Additionally, participants were asked to indicate how much they were satisfied with their experience using the AI tool on a 5-point Likert scale (1 = Very Dissatisfied to 5 = Very Satisfied).

*3.3.3. Behavioral Metrics*

Based on the event logs that were automatically captured within the AI platform, we collected several quantified measures that indicated participants' behavioral engagement during their interactions with the AI tool. These measures include total time spent (in seconds), total user message count, total user audio input message count, total user message characters, average user message characters.

*3.3.4. Academic Performance Scores*

As an indicator of learning outcomes, we collected participants' academic performance data, including their quiz scores from Units 2–5, test scores from Units 2–4, and their final exam scores. Quizzes consisted of multiple-choice and fill-in-the-blank questions, modeled after those found in the lessons, practice exercises, and problem sets. In contrast, tests required students to solve programming problems similar to those encountered during the course. Each test focused on the content of the corresponding unit but also built on knowledge from previous units. The Unit 5 test, serving as the final exam, was cumulative across the entire course. All quizzes and tests were timed and proctored using Honorlock.

**3.4. Data Analysis**

We used a mixed-methods approach to analyze the data. Quantitative analyses were conducted using SPSS Statistics (version 29), including descriptive statistics, multiple regression, and difference-in-differences (DiD) analysis. The regression model estimated the contribution of participants' demographic, engagement, UX, and interaction behavior variables to predicting perceived learning scores. The DiD approach compared test performance over time between the treatment and comparison groups to estimate the causal effect of the AI intervention while controlling for baseline performance. Qualitatively, thematic analysis of 141 open-ended responses was conducted to assess perceived changes in problem-solving, critical thinking, self-reflection, and verbal communication. Responses were inductively coded into themes indicating positive, no, or mixed change.

**4. Results**

**4.1. Patterns of Student Engagement and UX**

We first analyzed participants' behavioral patterns when interacting with the GenAI-powered assessment tool. On average, participants produced 83.6 messages across the entire assignment, with counts ranging from 32 to 237. Approximately 30% of all messages were submitted via audio input. The median[1] total completion time was 4,321 seconds (approximately 72 minutes). The mean total number of user message characters was 13,340.6 (roughly 2,668 words), with an average of 185.4 characters per message (about 37 words).

At the group level, the Unit 3 group generated a greater total number of user message characters (M = 10,993.4) than the Unit 4 group (M = 10,082.4), $t(169) = 4.93$, $p < .001$. This difference was expected, as the Unit 3 assignment contained more questions. When comparing message-level activity (per question), Unit 3 participants still produced more messages (M = 8.9) than Unit 4 participants (M = 7.9), $t(169) = 2.90$, $p < .01$. However, no statistically significant difference emerged in average message length between Unit 3 (M = 196.6) and Unit 4 (M = 175.3). Similarly, the average number of audio input messages per question was comparable between the two groups (Unit 3: M = 2.2; Unit 4: M = 2.8). Appendix A provides an illustrative example of a student–AI conversation using an excerpt from the actual interaction log.

---

[1] There were a few cases (about 10% of the total participants) where time tracking was either not feasible due to browser-related issues or inflated by unusually long session durations. These outliers skewed the mean; and therefore, median values were reported to more accurately represent the overall trend.

Participants generally reported high levels of engagement and positive UX with the AI platform (see Table 1). Behavioral engagement (BE) was the highest among the three engagement dimensions (M = 4.07, SD = 0.61), followed closely by affective engagement (AE) (M = 3.96, SD = 0.76) and cognitive engagement (CE) (M = 3.96, SD = 0.68), indicating that students were consistently active and emotionally involved while using the tool. In terms of UX, the participants rated the effectiveness of the platform most positively (M = 4.09, SD = 1.08), followed by usability (M = 3.99, SD = 1.16) and overall satisfaction (M = 3.88, SD = 1.23). These results suggest that the participants in general perceived the tool as usable and effective in promoting active engagement in learning.

Across most engagement and UX subscales, there were no statistically significant differences between the Unit 3 and Unit 4 groups. In other words, students' levels of engagement and perceptions of the tool were generally consistent regardless of the assignment unit. The only exception was AE, where students in Unit 4 reported significantly higher scores (M = 4.07, SD = 0.67) compared to those in Unit 3 (M = 3.83, SD = 0.84), $t(168) = -2.07$, $p = .04$. This finding suggests that students may have experienced a greater sense of emotional connection or interest during the Unit 4 assessment experience.

Table 1. Mean (Standard deviation) of engagement and UX sub-scales and *t*-test results for group comparison

|  | All Groups (n = 171) | Group Comparison | | df | t | p | Cohen's d |
| --- | --- | --- | --- | --- | --- | --- | --- |
|  |  | Unit 3 (n = 81) | Unit 4 (n = 90) |  |  |  |  |
| Engagement Sub-Scales |  |  |  |  |  |  |  |
| Affective | 3.96 (0.76) | 3.83 (0.84) | 4.07 (0.67) | 168 | -2.07 | .04 | -0.32 |
| Behavioral | 4.07 (0.61) | 4.03 (0.65) | 4.11 (0.57) | 168 | -0.80 | .43 | -0.36 |
| Cognitive | 3.96 (0.68) | 3.86 (0.73) | 4.05 (0.61) | 169 | -1.87 | .06 | -0.29 |
| UX Sub-Scales |  |  |  |  |  |  |  |
| Usability | 3.99 (1.16) | 4.04 (0.65) | 3.94 (1.39) | 169 | 0.53 | .59 | 0.08 |
| Effectiveness | 4.09 (1.08) | 4.17 (0.73) | 4.02 (1.32) | 169 | 0.91 | .37 | 0.14 |
| Overall Satisfaction | 3.88 (1.23) | 3.89 (1.05) | 3.88 (1.38) | 169 | 0.06 | .95 | 0.01 |

Correlation analyses revealed several significant patterns among the participants' UX, engagement, and interaction behaviors while using the tool (see Table 2). First, all UX subscales were positively and significantly correlated with all three engagement subscales with $r$s ranging from .33 to .54, $p$s < .01. This indicates that the more students found the tool more usable and effective, the more likely they were to feel emotionally invested, actively participate, and engage cognitively during the AI-mediated assessments. In terms of the participants' actual behavior during the assignments, we observed that their average user message character count had small but significant, positive correlations with AE ($r = .17$, $p < .05$) and CE ($r = .18$, $p < .05$). Interestingly, the user audio input count variable was also significantly and positively correlated with all three engagement sub-scales: CE ($r = .20$, $p < .01$), AE ($r = .26$, $p < .01$), and BE ($r = .22$, $p < .01$). These findings suggest that students' active participation in Socratic dialogue, whether through verbal or written responses, is closely linked to increases in their emotional, behavioral, and cognitive engagement. Conversely, total time spent on the platform was negatively associated with CE ($r = -.18$, $p < .05$), implying that longer durations could be associated with cognitively shallow involvement.

Table 2. Bivariate correlations among the engagement, UX, and interaction behavior variables

|  | 1 | 2 | 3 | 4 | 5 | 6 | 7 | 8 | 9 | 10 | 11 | 12 | 13 |
| --- | --- | --- | --- | --- | --- | --- | --- | --- | --- | --- | --- | --- | --- |
| AE (1) | 1.00 |  |  |  |  |  |  |  |  |  |  |  |  |
| BE (2) | .72** | 1.00 |  |  |  |  |  |  |  |  |  |  |  |
| CE (3) | .62** | .64** | 1.00 |  |  |  |  |  |  |  |  |  |  |
| Usability (4) | .39** | .44** | .33** | 1.00 |  |  |  |  |  |  |  |  |  |

| | | | | | | | | | | | | |
|---|---|---|---|---|---|---|---|---|---|---|---|---|
| Effectiveness (5) | .41** | .50** | .35** | .88** | 1.00 | | | | | | | |
| Satisfaction (6) | .54** | .43** | .34** | .74** | .77** | 1.00 | | | | | | |
| Total Time Spent (7) | -.09 | -.11 | -.18* | -.10 | -.10 | -.07 | 1.00 | | | | | |
| Copied AI Message (8) | -.06 | -.06 | .03 | .01 | -.04 | -.06 | -.08 | 1.00 | | | | |
| Pasted Large Text (9) | -.14 | -.18* | -.08 | -.06 | -.08 | -.12 | .01 | .65** | 1.00 | | | |
| User Msg. Ct. (10) | -.03 | -.001 | -.14 | .004 | .05 | .05 | .20** | -.30** | -.08 | 1.00 | | |
| User Audio Input Ct. (11) | .26** | .22** | .20** | .10 | .07 | .13 | -.07 | -.25** | -.44** | .26** | 1.00 | |
| Total User Msg. Chars. (12) | .17* | .14 | .14 | .14 | .11 | .13 | .004 | .33** | .51** | -.11 | .05 | 1.00 |
| Avg User Msg. Chars. (13) | .17* | .14 | .18* | .14 | .12 | .12 | -.05 | .32** | .42** | -.44** | -.07 | .87** | 1.00 |

*Note.* * indicates $p < .05$, ** $p < .01$, *** $p < .001$.

### 4.2. Impact of the AI-Powered Assessment Tool on Perceived and Actual Learning Outcomes

#### *4.2.1. Perceived Learning Outcomes*

We first examined the correlations between participants' perceived learning (PL) scores and their levels of engagement, UX, and interaction behaviors during the AI-based assignments. As expected, PL was strongly and positively correlated with all three engagement sub-scales: AE ($r = .76, p < .01$), BE ($r = .66, p < .01$), and CE ($r = .58, p < .01$). These high correlations suggest that feeling emotionally connected, behaviorally invested, and cognitively simulated during the assessment activity was associated with perceiving greater learning benefits. Similarly, PL showed moderate, significant correlations with all three UX sub-scales: usability ($r = .45, p < .01$), effectiveness ($r = .50, p < .01$), and satisfaction ($r = .54, p < .01$). These findings indicate that students who found the tool easy to use, effective in supporting learning, and satisfying were more likely to report stronger perceived learning outcomes. Among the interaction behaviors variables, three indicators of participation had small but significant, positive correlations with PL: user audio input message count ($r = .16, p < .05$), total user message character count ($r = .17, p < .05$), and average user message character count ($r = .21, p < .05$).

     To examine the relationship between participants' perceived and actual learning outcomes, we conducted a correlation analysis between PL scores and assessment scores obtained immediately following the treatment. Although most correlations were not statistically significant, within the Unit 3 group we observed significant negative correlations between PL scores and both the Unit 3 quiz ($r = -.30, p < .01$) and test scores ($r = -.35, p < .01$). These findings point to a potential misalignment between students' perceived and actual learning when engaging with the AI tool.

     Next, a multiple regression analysis was conducted to identify which factors significantly predicted the PL scores (see Table 3). The overall model was statistically significant, $F (18, 142) = 17.19, p < .001$, and explained approximately 69% of the variance in PL scores. Among all predictors, engagement subscales emerged as the strongest contributors, particularly AE ($\beta = .445, p < .001$) and BE ($\beta = .181, p < .05$), even after accounting for demographic factors, UX variables, and interaction behavior patterns. In other words, feeling interested, motivated, and emotionally invested, as well as engaging consistently in learning activities, appears to have a positive impact on perceived learning outcomes when using the GenAI-assisted learning tool. Additionally, English as a native language was a significant positive predictor ($\beta = .156, p < .01$), indicating that participants who were native English speakers tended to report greater perceived learning. Also, identifying as male negatively predicted the PL scores ($\beta = -.116, p < .05$), suggesting potential gender-related differences in perceived learning. On the other hand, none of the interaction behavior variables significantly predicted PL scores.

Table 3. Results of multiple regression analysis predicting perceived learning scores

| Variable | B | SE B | β |
|---|---|---|---|
| Condition | 0.110 | 0.088 | 0.081 |
| Male | -0.172 | 0.074 | -0.116* |
| Age | 0.056 | 0.034 | 0.084 |
| USA | -0.146 | 0.094 | -0.087 |
| English | 0.288 | 0.114 | 0.137* |
| White | -0.147 | 0.074 | -0.108* |
| Usage of GenAI | -0.001 | 0.032 | -0.002 |
| AE | 0.395 | 0.071 | 0.445*** |
| BE | 0.203 | 0.088 | 0.181* |
| CE | 0.099 | 0.066 | 0.099 |
| Usability | -0.030 | 0.062 | -0.050 |
| Effectiveness | 0.089 | 0.072 | 0.138 |
| Satisfaction | 0.069 | 0.048 | 0.122 |
| Total Time Spent | 0.000 | 0.000 | 0.022 |
| Total Message Character Count | 0.000 | 0.000 | -0.090 |
| Average Message Count per Q | 0.027 | 0.020 | 0.094 |
| Average Message Characters Count | 0.001 | 0.001 | 0.181 |
| User Audio Input Message Count | -0.001 | 0.001 | -0.034 |
| (Constant) | -0.811 | 0.769 | |
| $R^2$ | | 0.685 | |
| F | | 17.193*** | |

*Note.* * indicates $p < .05$, ** $p < .01$, *** $p < .001$.

### 4.2.2. Actual Academic Performance Outcomes

When the unit-level assessment results were compared between the Unit 3 and Unit 4 groups, their academic performances were generally similar, as illustrated in Figure 2. Baseline performance measures obtained from the Unit 2 quiz and test, conducted before any interaction with the AI tool, also did not differ significantly between the groups. This finding suggests that both groups had similar levels of prior knowledge. We observed that the Unit 3 group achieved slightly higher scores on the Unit 3 quiz and test compared to the Unit 4 group. However, these differences were descriptive and not statistically significant, indicating comparable performance levels across the two groups immediately following their respective AI interventions. Moreover, results on the subsequent unit assessments (Unit 4 and Unit 5) and final course scores showed high consistency between the groups.

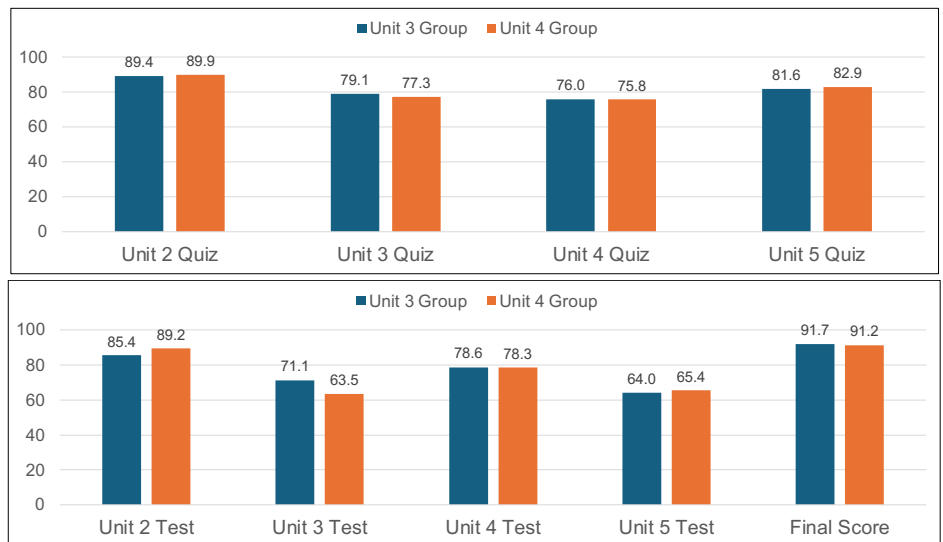

**Fig. 2** Comparison of unit-level assessment results and final scores between the groups

Next, using a DiD approach, we conducted a multiple regression analysis to further examine the effect of engagement with the AI-driven assessment tool on participants' academic performance over time, specifically comparing scores before and after the assignment intervention. The regression model included baseline scores measured prior to the assignment, assignment condition (Unit 3 or Unit 4), time relative to treatment (pre- or post-assignment), and the interaction term capturing the treatment effect. The overall model was statistically significant with $F(4, 793) = 86.02$, $p < .001$ for the quiz model and $F(4, 793) = 61.11$, $p < .001$ for test model, accounting for approximately 30% of the variance in quiz scores and 24% in test scores (see detailed results in Table 4).

As expected, quiz performance declined in the later units, which is consistent with the increasing difficulty of the course content in Units 3 and 4. However, a significant positive interaction effect emerged ($\beta = .107$, $p < .05$), suggesting that engagement with the assessment tool mitigated this performance decline. While students overall experienced a decrease of approximately 11.84 points on average from pre- to post-intervention, those who participated in the AI-based assignment saw this drop reduced by an estimated 3.30 points (see Figure 3). In other words, students who participated in the assignment exhibited a less pronounced drop in their quiz scores compared to their peers, despite the increasing difficulty of the course content. However, this buffering effect was not observed in the unit test scores ($\beta = -.053$, $p > .05$). This was somewhat expected, considering that the questions in the AI-based assignment were designed based on the quiz topics.

Table 4. Results of difference-in-differences multiple regression analysis: Quiz as a dependent variable

| Variable | Quiz Model | | | Test Model | | |
| --- | --- | --- | --- | --- | --- | --- |
| | B | SE B | β | B | SE B | β |
| Condition | -4.678 | 0.978 | -0.160*** | -3.368 | 1.884 | -0.063 |
| Post Scores | -11.836 | 1.373 | -0.406*** | -3.300 | 2.631 | -0.062 |
| Pre Scores | 0.749 | 0.049 | 0.453*** | 0.634 | 0.042 | 0.477*** |
| Post*Treatment | 3.300 | 1.373 | 0.107* | -3.017 | 2.631 | -0.053 |
| (Constant) | 21.446 | 4.475 | - | 27.568 | 3.950 | - |
| F | | 86.018*** | | | 61.105*** | |
| $R^2$ | | 0.303 | | | 0.236 | |

*Note.* Post scores for Unit 3 group reflect scores from Unit 3 and Unit 4 assessments, while those for the Unit 4 group reflect scores from Unit 4 assessments. Pre scores for Unit 3 group reflect scores from Unit 2; while those for Unit 4 group reflect scores from Unit 2 and Unit 3 assessments. * indicates $p < .05$, ** $p < .01$, *** $p < .001$.

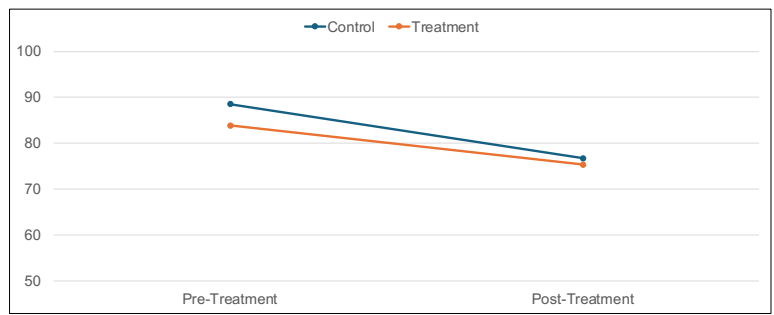

**Fig. 3** Buffering effect of the GenAI-powered tool on quiz scores

To further isolate the effect of the AI intervention while controlling pre-intervention performance, an extended multiple regression analysis was conducted. In addition to the experimental condition, pre-and post-intervention quiz scores, and the interaction capturing the treatment effect, this model included a treatment-by-pre quiz scores interaction term. The overall model was statistically significant, $F(5, 792) = 69.99$, p < .001, and accounted for 31% of the variance in quiz scores ($R^2 = .306$) (see Table 5). As expected, prior quiz performance strongly predicted future quiz performance (β = .497, *p* < .001), validating that students' baseline academic ability played a major role in shaping later outcomes. We also observed a significant positive interaction between treatment and post quiz scores (β = 0.736, *p* < .05), which indicates a meaningful buffer against the general decline observed in post-assignment quiz scores. Notably, there was a significant negative interaction between pre-intervention scores and treatment (β = –.635, *p* < .05). Based on the unstandardized coefficient (B = –0.217), for every 1-point lower baseline score, the benefit of the assessment tool increased by approximately 0.21 points. This finding suggests that students with initially lower prior performance may have experienced greater gains from engaging with the AI-powered personalized activity.

Table 5. Results of extended difference-in-differences multiple regression analysis: Quiz as a dependent variable

| Variable | B | SE B | B |
|---|---|---|---|
| Condition | -4.670 | 0.976 | -0.160*** |
| Post Scores | -11.813 | 1.370 | -0.495*** |
| Pre Scores | 0.822 | 0.060 | 0.497*** |
| Post*Treatment | 22.764 | 9.393 | 0.736* |
| Pre*Treatment | -0.217 | 0.104 | -0.635* |
| (Constant) | 14.942 | 5.439 | - |
| F | | 69.986*** | |
| $R^2$ | | 0.306 | |

*Note*. * indicates *p* < .05, ** *p* < .01, *** *p* < .001.

### 4.3. The Influence of Socratic Questioning on Higher-Order Thinking Skills

By conducting a thematic analysis of participants' short open-ended survey responses (*n* = 141), we examined whether students perceived improvements in four specific types of higher-order thinking skills after interacting with the AI tool. These skills include problem-solving, critical thinking, self-reflection, and verbal communication. Students generally reported the strongest gains in problem-solving, followed by notable perceived improvements in critical thinking and self-reflection, and relatively limited changes in verbal communication, primarily due to variations in using the verbal interaction feature. Figure 4 compares the participants' perceived influence of the tool on their higher-order thinking skills, and Appendix B provides detailed results with representative quotes from participants.

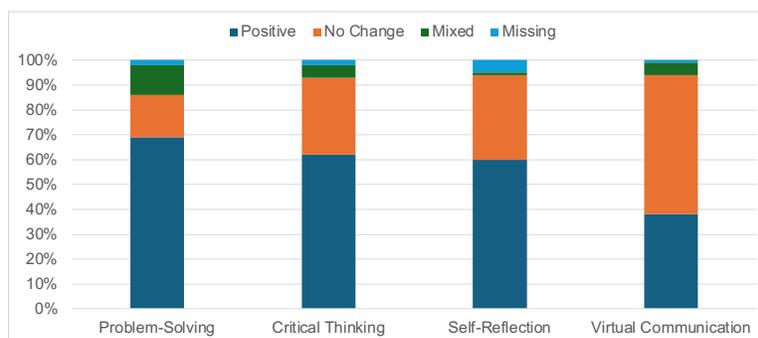

**Fig. 4** Comparison of participants' responses in percentage regarding their perceived influence of the GenAI-powered assessment tool on their higher-order thinking skills

The majority of the survey respondents (69%) reported experiencing positive changes in problem-solving and often described improvements such as more structured step-by-step reasoning, deeper conceptual understanding of coding tasks, and greater confidence in debugging practices. The participants frequently attributed these enhancements to the AI's Socratic questioning method, which prompted them to think critically, make connections across multiple concepts, and clearly articulate their thought processes—skills essential for computational thinking within computer science education. In contrast, approximately 17% of respondents reported no noticeable changes, often citing the brief duration of engagement with the tool as a limiting factor. An additional 12% stated that although their overall problem-solving approach had not significantly changed within this brief exposure, they still found value in the activity, such as reinforcing existing knowledge, promoting metacognitive awareness, or revealing gaps in their understanding.

Similarly, a significant proportion of participants (62%) reported positive developments in their critical thinking skills. Students specifically highlighted improvements in their ability to think more analytically, explain reasoning more clearly, and evaluate coding logic from multiple perspectives. They noted that the structured questioning provided by the AI tool encouraged them to slow down, question assumptions, and explore alternative approaches to solving problems. Conversely, 31% reported no meaningful change in their critical thinking, commonly noting limited time or interactions with the tool as possible reasons. Additionally, 5% expressed mixed experiences; while they did not observe a distinct change in critical thinking per se, they acknowledged secondary benefits, such as improved clarity of coding concepts or better comprehension of programming logic.

In terms of self-reflection, 60% of respondents reported perceived gains in their reflective thinking skills after engaging with the tool. These participants generally mentioned becoming more deliberate and thorough in evaluating their own understanding and thought processes. Some of them noted that the tool prompted them to ask follow-up questions, compare alternatives, and articulate their logic more clearly. Others mentioned that this iterative and reflective dialogue contributed to fostering intentional learning behaviors and metacognitive awareness. In contrast, 34% explicitly reported no change in their self-reflection skills mainly due to limited exposure to the tool or insufficient time to observe meaningful effects. About 1% provided mixed feedback, such as perceiving indirect benefits while remaining uncertain about their perceived development in self-reflection.

Finally, we observed that only 38% of respondents reported positive changes in verbal communication skills. These participants frequently noted improved clarity when explaining coding logic, enhanced precision with technical terminology, and increased comfort in communicating computing concepts. However, more than a half of the participants (56%) indicated no perceived change in verbal communication, largely due to their decision to type responses rather than utilize the AI's verbal interaction feature. They often cited technical issues or environmental constraints. Additionally, 5% described mixed experiences, such as increased awareness of how to explain concepts clearly or improved thinking-aloud strategies, even if they did not perceive direct improvements in their overall verbal skills.

## 5. Discussion

This study examined the impact of a GenAI-powered assessment tool grounded in Socratic questioning on student engagement and learning outcomes in a large-scale introductory computer science online course. Rather than providing answers directly, the tool engaged students in dialogue that prompted them to explain, justify, and reflect on their reasoning. For example, when students were asked to compare loop control structures and apply them to a real-world problem, the tool guided them through successive stages of reasoning. It first prompted them to articulate differences between *for* and *while* loops, then pushed them to construct pseudocode showing how their chosen loop would work in practice. A second example comes from a debugging exercise in which students identified and corrected code errors. Here, the tool asked not only what the error was, but also why it occurred. For instance, after a student attempted to fix a function that manipulated strings, the agent might ask: "Exactly right! And why do we get this specific error message? What property of strings in Python does this error point to?" Such follow-up questions provided students with opportunities to solidify their conceptual understanding. Overall, our study contributes to the growing literature on GenAI's promising role in fostering deep engagement, personalized learning, and scaffolding when used interactively (Gu & Yan, 2024; Sajja et al., 2025; Yan et al., 2024).

Our findings offer insight into student engagement by drawing on self-reported perceptions and system-generated interaction logs. The results highlight that the GenAI tool consistently provided students with engaging learning experiences, as indicated by high levels of engagement scores across all three engagement dimensions and across the two treatment groups. Notably, markers of active and authentic participation, such as providing elaborated and lengthy responses were significantly correlated with reporting higher levels of emotional and cognitive engagement. Additionally, our findings indicate that students' levels of engagement with the tool were closely associated with their perceived UX in terms of both the interface usability and effectiveness of the interactive AI features. These patterns echo prior work on AI-mediated dialogue and underscore the need for careful interaction design to promote authentic engagement (Fakour & Imani, 2025; Yan et al., 2024). Our findings emphasize the importance of providing students with a straightforward, easy-to-navigate interface as well as personalized feedback that aligns with each student's level of understanding (Authors, 2025; Kohler, 2022; Peters et al., 2018; Sajja et al., 2025).

A particularly notable finding was the deeper engagement and cognitive effort demonstrated by students who used the tool via audio input instead of typing. This pattern suggests that incorporating oral interaction modes into AI-powered assessment can add value to formative assessment practices. Oral assessments require students to verbalize reasoning in real time and offer opportunities for self-reflection and identification of misconceptions or knowledge gaps (Qi et al., 2023; Reckinger & Reckinger, 2022; Theobold, 2021). In that regard, AI-driven oral assessments can offer rich diagnostic insights for educators that can enable timely and tailored instructional interventions (Gardner & Giordano, 2023). Future research should explore how this AI platform and other similar AI-powered assessment tools can facilitate reliable oral interaction while ensuring accessibility and effective integration across diverse learning environments.

Our regression analyses provided strong evidence that active interaction with the GenAI tool was significantly associated with students' perceived learning outcomes, even after controlling for the UX variables, demographics, and interaction patterns. Students who felt more emotionally and behaviorally engaged consistently reported greater learning benefits, which is consistent with prior research findings (Gu & Yan, 2024; Yilmaz & Yilmaz, 2023). It is noteworthy that certain demographic variables, specifically gender and English as a native language, also emerged as significant predictors of perceived learning, suggesting that students' backgrounds may influence how they experience and benefit from AI-supported learning environments. Future research should further explore how demographic characteristics may moderate the impact of AI engagement on learning outcomes, and whether tailored interventions can help close potential gaps in perceived or actual learning benefits across diverse student populations.

Furthermore, our quasi-experimental study yields empirical evidence of the tool's positive learning impact and therefore contributes to extending the literature on GenAI in Education (Bauer et al., 2025). Beyond analyzing students' self-reported learning outcomes, we evaluated the learning effects using students' actual academic performance data. Using a DiD approach, our study revealed that as the course content became more difficult, students in the treatment group experienced a smaller decline in quiz scores, approximately 3.3 points less than peers who did not use the tool. This buffering effect was particularly noticeable among lower-performing students, suggesting that GenAI-supported Socratic dialogue can help close performance gaps in challenging content areas. These results align with previous research demonstrating the potential value of Socratic questioning in challenging student to articulate ideas and reasoning and thereby cultivating scientific inquiry and critical thinking (Fakour & Imani, 2025; Yang et al., 2005). While the quizzes primarily consisted of multiple-choice and fill-in-the-blank items, the tests focused on open-ended programming problems. Given that our GenAI-powered tool was designed to support conceptual reasoning and higher-order thinking, rather than direct test preparation or coding practice, we anticipated stronger performance improvements on quizzes than on tests.

Interestingly, despite students' widespread perception that the tool supported their learning, their perceived learning did not consistently align with actual quiz or test performance. This discrepancy likely reflects a combination of factors. One possibility is retrospective bias, as participants reported their PL scores at the end of the semester rather than immediately following the treatment. In addition, students may have overestimated their learning gains due to the generally positive experience of interacting with the GenAI tool. Importantly, the tool was intentionally designed to foster conceptual understanding and metacognitive reflection rather than to directly prepare students for conventional exams. The assignment questions were not meant to mirror the format or difficulty of quiz or test items; instead, they served as formative assessments to prompt reasoning, address conceptual misunderstandings, and encourage reflective thinking. This design choice may help explain the misalignment between perceived and actual outcomes. Moving forward, future research should develop more robust measures to capture students' gains in conceptual understanding and higher-order thinking, which would provide a more accurate evaluation of the educational impact of AI-assisted learning tools.

The findings of our work also make a timely contribution to the ongoing debate on whether GenAI tools hinder or enhance students' higher-order thinking skills (e.g., Kosmyna et al., 2025; Stadler et al., 2024). In our study, a substantial proportion of the participants reported gains in problem-solving, critical thinking, and self-reflection as a result of interacting with the tool. Qualitative feedback further suggested that students found the structured, Socratic-style interactions as beneficial for moving beyond surface-level understanding, developing analytical thinking and clarifying logic in programming tasks. Additionally, these benefits were especially salient in responses describing how the tool encouraged students to ask themselves better questions, consider alternatives, and challenge assumptions. These findings align with earlier studies showing that when AI is used as a scaffold, it can enhance critical thinking rather than inhibit it (Fakour & Imani, 2025; Favero et al., 2024).

Moreover, our findings offer a robust proof-of-concept for leveraging GenAI technologies to develop a formative assessment tool that can empower both students and instructors. Participants' reflections on perceived gains in reflective thinking skills suggest that engaging with AI-powered Socratic dialogue created opportunities for students to actively monitor and evaluate their own understanding with personalized feedback and thinking processes based on personalized feedback. By identifying misconceptions and offering step-by-step guidance toward conceptual clarity, the tool functioned as a proactive scaffold for individual learning. Such outcomes align with prior research that has demonstrated the potential of GenAI technologies to foster metacognitive awareness and self-directed learning (Weng et al. 2024; Yan et al. 2024).

Importantly, unlike commercially available GenAI platforms such as OpenAI's *study mode* or Anthropic's *learning mode*, this tool grants instructors meaningful pedagogical agency by allowing them to upload lecture-specific materials, design rubric-based questions and guided answers, and intentionally shape students' learning trajectories. This educator-driven design grounded on the human-AI partnership ensures that GenAI-facilitated feedback is both contextually relevant and pedagogically meaningful while addressing concerns about generic or uncontextualized AI outputs (Luo et al., 2024). By combining AI's adaptive capabilities with intentional instructional design, the tool fosters not only deeper cognitive engagement but also a stronger alignment between technology and pedagogy.

## 6. Limitations

While the findings offer valuable insights into the impact of GenAI-powered Socratic dialogue in computer science education, several limitations must be acknowledged. The study was conducted in a single undergraduate computing course at one institution, which may limit the generalizability of the results to other disciplines, educational contexts, or student populations. Second, while we employed rigorous statistical controls, the quasi-experimental design cannot fully rule out selection bias or unmeasured confounding variables such as students' prior programming experience, intrinsic motivation, or external support systems. Although we measured behavioral engagement quantitatively, we lacked qualitative data such as interviews or think-aloud protocols that could illuminate students' thought processes during AI interaction. Additionally, the short duration between the intervention and outcome assessments may not capture the long-term effects of the GenAI tool on conceptual understanding or retention.

Finally, while oral interaction appeared to enhance cognitive engagement, several barriers emerged. Some students experienced technical issues or preferred text-based formats due to comfort or environmental constraints. In the context of CS 1301 specifically, the nature of the course content posed additional limitations: many questions in the AI tool involved programming tasks that are not easily or efficiently addressed through audio input. As such, it is reasonable that students would rely more on text-based responses when interacting with the AI tool. These findings suggest that the success of AI-driven interventions depends not only on pedagogical design but also on usability, technical reliability, accessibility, and content compatibility. Also, future studies need to explore whether longer-term implementation across multiple assignments or courses may allow students to more fully develop their verbal communication and metacognitive skills.

**7. Conclusion**

Our study demonstrates that Socratic Mind, the GenAI-powered assessment tool effectively promoted meaningful engagement and supported learning in the context of introductory computer science education in higher education. Students reported high levels of affective, behavioral, and cognitive engagement, and generally perceived the tool as usable and beneficial. Perceptions of learning were strongly associated with engagement and UX. Notably, students who engaged with the tool experienced a significantly smaller decline in quiz scores over time, especially those with lower prior achievement, suggesting a buffering effect that helped struggling learners. Many students noted that using the tool helped them improve their problem-solving abilities and critical thinking. Beyond these empirical findings, our study extends work on dialogic learning by illustrating how Socratic questioning can be operationalized at scale through GenAI in a large, online course. For educators, the results suggest that such tools can complement—not replace—traditional assessments by providing personalized, dialogic feedback that encourages reflection and reasoning. Overall, as institutions explore GenAI integration in education, this study shows that a dialogic, AI-powered assessment tool can offer a scalable strategy to promote students' deep engagement and higher-order thinking.

**Appendix A. Illustration of Student Interaction with the GenAI-powered Assessment Tool**

Figure A1 shows that the GenAI-powered assessment tool's messages in the dark blue text box and a student's audio input messages transcribed in the orange text box. During this interaction, the student spent a total of 4 minutes, 54 seconds.

> Hey! I heard you're learning about Python exception handling in your computing class. Can you tell me what it's all about? I'm really curious!

> Okay, exception handling. It basically helps you catch errors in your code and allow it to keep going instead of crashing. You can use the try function to try a block of code if an error occurs. You can use the accept function to do that. You can also use finally after all that to make that code run under finally no matter what. You can also, if the block of code in try doesn't yield an error after the accept, you can put an else and that will run, but only if try doesn't run an error.

> Thanks for explaining! I think I understand the basic idea - it's about preventing crashes when errors happen. I have a few questions though:
> You mentioned using the "accept" function to catch errors. Did you mean "except"? I've been looking at some Python code examples online and they seem to use "except" rather than "accept".
> Also, could you give me a real-world example of why we'd need exception handling? Like, when would our code crash without it?

> I did mean except not accept and a real-world example of when we need exception handling so if you have If you have code that you go if, or sorry, for i in range 1 to, or from 10 to negative 10 and then you divide 10 by I and then you have I equals I minus 1 and then have that keep running eventually you're going to run into 10 divided by 0 which is an error and so which is an error in your code which would cause it to crash unless you used exception handling you could catch the error and continue your code

I see what you're saying! So you're describing a scenario where we're dividing by a variable that eventually becomes zero, which would cause a division by zero error.
Let me make sure I understand your example correctly. You're saying something like:

```
for i in range(10, -10, -1):   # Count from 10 down to -9
    result = 10 / i            # This will cause an error when i beco
    print(result)
```

Without exception handling, this would crash when i becomes 0, right? And with exception handling, we could catch that specific error and keep the program running.

Is that what you meant? And just to double-check, it's "except" (not "accept") that we use to catch the exceptions, right?

Yes, that is what I meant.

Great! That's a perfect example of when exception handling is useful. When we're dealing with user input, file operations, or calculations like your division example, things can go wrong that would normally crash our program. So the basic structure would be something like:

```
try:
    # Code that might cause an error
    for i in range(10, -10, -1):
        result = 10 / i
        print(result)
except ZeroDivisionError:
    # Code to handle the specific error
    print("Cannot divide by zero!")
else:
    # This runs if no exceptions occurred in the try block
    print("No errors occurred!")
finally:
    # This code always runs, whether there was an exception or not
    print("Calculation completed.")
```

Is that how you understand it too? The "try" block contains risky code, "except" catches specific errors, "else" runs only if no errors occurred, and "finally" runs no matter what.

> Yep, that's my understanding.

Awesome! You've got a solid grasp of exception handling in Python. Just to summarize what we've covered:
1. Exception handling helps prevent program crashes when errors occur
2. The basic structure uses **try**, **except**, **else**, and **finally** blocks
3. **try** contains code that might cause errors
4. **except** catches specific errors (like ZeroDivisionError)
5. **else** runs only if no errors occurred in the try block
6. **finally** runs regardless of whether an error occurred

Your division example was perfect - it shows a real situation where exception handling is necessary. Without it, dividing by zero would crash the program, but with exception handling, we can gracefully handle the error and continue execution.

Is there anything else about exception handling you're curious about?

> Yes, with the accept function you can catch specific types of errors like the serial division error, type error, name error, syntax error, or you can also just do exception which will catch all errors, which is interesting.

You're absolutely right! That's an important aspect of exception handling. Just to clarify again, it's "except" rather than "accept" in Python, but you've got the concept correct. You can indeed catch specific types of errors or use a general exception handler. Here's how that might look:

```python
try:
    # Risky code here
    result = 10 / 0
    print(undefined_variable)  # This would cause a NameError
except ZeroDivisionError:
    print("Can't divide by zero!")
except NameError:
    print("That variable doesn't exist!")
except Exception as e:
    print(f"Some other error occurred: {e}")
```

The order is important too - you should catch specific exceptions first, then more general ones. If you put the general **except Exception** first, it would catch everything and the more specific handlers would never run. One thing to note though - catching all exceptions with **except Exception** can sometimes hide bugs in your code. It's usually better to catch only the specific exceptions you expect and know how to handle. What do you think about that?

> I think that's how I understand it.

> Great! You've got a solid understanding of Python exception handling. It's a really useful tool that will help make your programs more robust.
> To recap our conversation:
> - Exception handling prevents program crashes when errors occur
> - The structure uses **try**, **except**, **else**, and **finally** blocks
> - You can catch specific exceptions (like ZeroDivisionError, TypeError, NameError)
> - You can use a general **except Exception** to catch all errors, but it's better practice to catch specific ones when possible
> - Your division by zero example showed a practical use case for exception handling
>
> This will definitely come in handy as you continue learning Python. Good luck with your studies!

Session Summary: The student successfully explained Python exception handling, covering the try-except-else-finally structure and demonstrating understanding of how to catch specific error types. They provided a practical example of division by zero to illustrate why exception handling is necessary. The conversation clarified the correct terminology ("except" rather than "accept") and discussed best practices like catching specific exceptions rather than using general exception handlers.

**Fig. 5** An illustrated sample interaction from the Unit 3 Error Handling topic question.

**Appendix B. A Summary of Perceived Changes in Higher-Order Thinking Skills and Sample Quotes (n = 141)**

| Higher-Order Skills | Participant Response Percentage | | Sample Quotes |
|---|---|---|---|
| Problem-Solving | Positive Change | 69% | - *I improved my problem-solving skills and now focus on context clues given in the problems on homework and exams.*<br>- *Yes, the (blinded) made me carefully walk myself through the problems. As a result, I was able to, for example, debug my code faster and understand what was going on quickly.*<br>- *I was able to logically think through each question and explain my reasoning to each problem.* |
| | No Change | 17% | *I don't think it changed my problem-solving skills overall.* |
| | Mixed Experience | 12% | *Not really, but it helped me understand more about try-except blocks* |
| Critical Thinking | Positive Change | 62% | - *I think my critical thinking skills improved. (blinded) made me think more in depth about coding and why some concepts work the way they do, especially when it comes to lists and dictionaries.*<br>- *it made me critically think about how certain codes work using the rules I had been taught in the course.*<br>- *I noticed that I can translate what a block of code is trying to do because of (blinded). It made me evaluate what the code was doing and why.* |
| | No Change | 31% | - *I don't really think that after just one i've changed my critical thinking much, however, I could see the SM having more of an impact if we used it for the whole course*<br>- *I have not noticed any major changes* |
| | Mixed Experience | 5% | *I haven't noticed many changes, but I think I can understand the reasoning behind why I choose certain codes to approach certain problems. Like knowing why to use a for loop over a while loop.* |
| Self-Reflection | Positive Change | 60% | - *I definitely feel that I reflect more clearly on the code I am writing as a result of (blinded).*<br>- *I was more thorough with my understanding and was able to elaborate a lot more on certain concepts.*<br>- *I did find myself reflecting more after using (blinded), and doing so in more complex ways. For example, the program challenged me to think less linearly about topics when revisiting them to ensure my understanding is thorough.* |
| | No Change | 34% | - *I was not able to become that much more self critical.*<br>- *No major differences.* |
| | Mixed Experience | 1% | - *Not necessarily, I just feel more confident in my coding abilities.* |
| Verbal Communication | Positive Change | 38% | - *Having to verbally explain the questions improved my understanding of the concepts and helped me to improve how I can communicate my knowledge.*<br>- *I've gotten better at describing coding functions and verbally de-bugging code snippets with people.*<br>- *I learned that I had to be more precise in my wording and more specific. almost every time I* |
| | | | *would answer a question, I was told to be more specific.*<br>*- I found out I tried to talk more clearly and straight to the point when I was explaining the concepts, since that was easier to communicate my knowledge to the AI.* |
| | No Change | 56% | *- I didn't use the verbal feature; I typed my responses, and so I haven't noticed any verbal changes.*<br>*- I typed all of my responses so I don't feel a significant change in verbal communication skills.* |
| | Mixed Experience | 5% | *- I think my verbal communication skills haven't seen a major change but rather a slight improvement in that I definitely think about various ways to answer a question before just saying the usual comfortable answer.*<br>*- Not really, still felt like talking to a screen or AI, so I didn't communicate the way I normally would. If anything, I was more technical when explaining things, but I think that had more to do with how the program was structured. The more technical explanation of the content just made it easier to complete things faster.* |

*Note*. The percentage of missing responses was not included in the summary due to lack of relevance.